\documentclass[a4paper]{jpconf}
\usepackage{graphicx}
\begin{document}

\title{Atmospheric Neutrinos\footnote{Research
supported in part by the U.S. Department of Energy under DE-FG02 91ER40626.}
}
\author{Thomas K. Gaisser}
\address{
Bartol Research Institute and Department of Physics and Astronomy\\ 
University of Delaware,
Newark, DE 19716 USA}

\ead{gaisser@bartol.udel.edu}

\begin{abstract}
This paper is a brief overview of the theory and experimental data of atmospheric 
neutrino production at the fiftieth anniversary of the experimental
discovery of neutrinos.
\end{abstract}

\section{Introduction}

\begin{figure}[h]
\begin{minipage}{18pc}
\includegraphics[width=18pc]{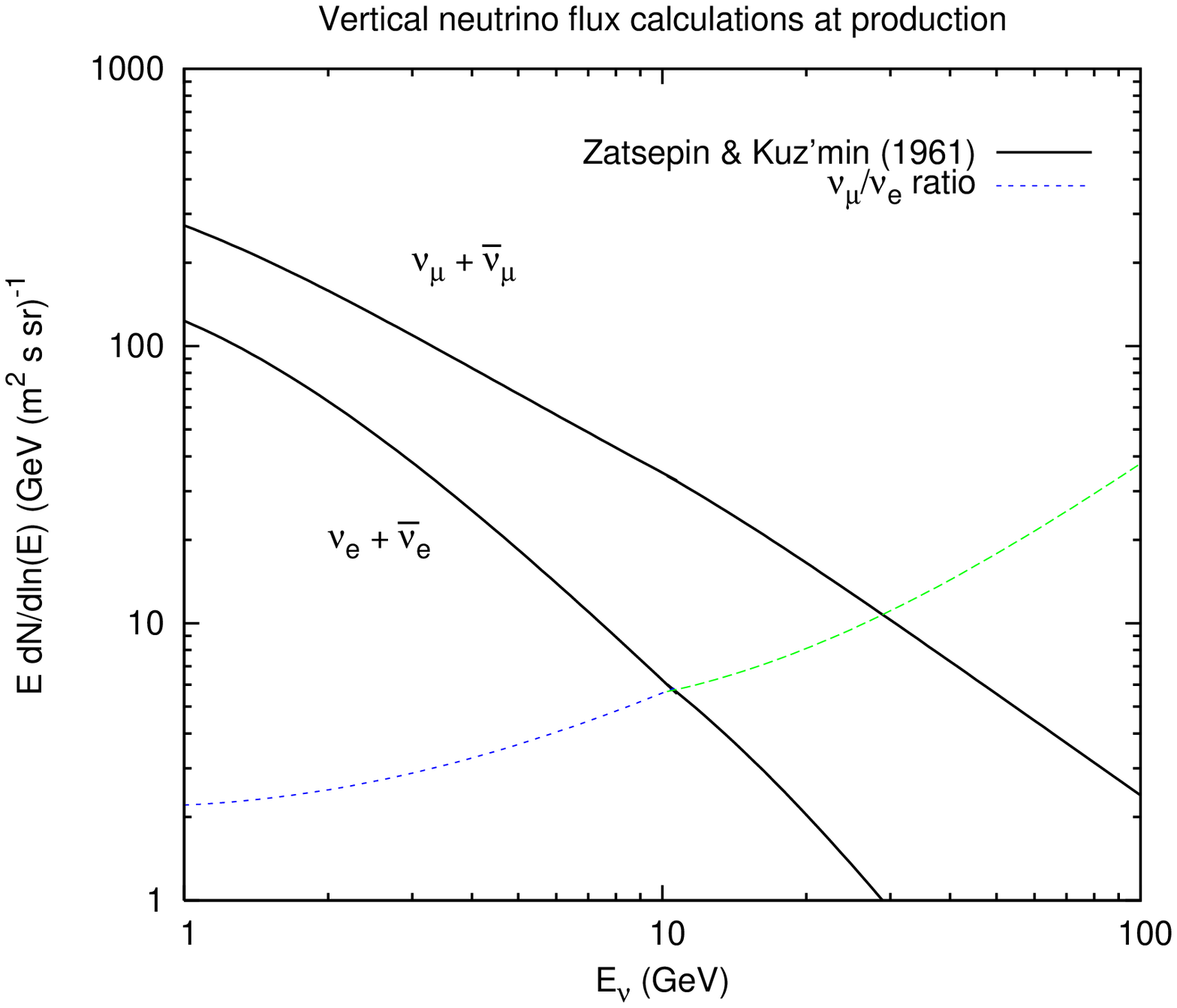}
\caption{\label{fig1a}Plots of the numerical formulas of
Ref.~\protect\cite{ZK2}.}
\end{minipage}\hspace{2pc}%
\begin{minipage}{18pc}
\includegraphics[width=18pc]{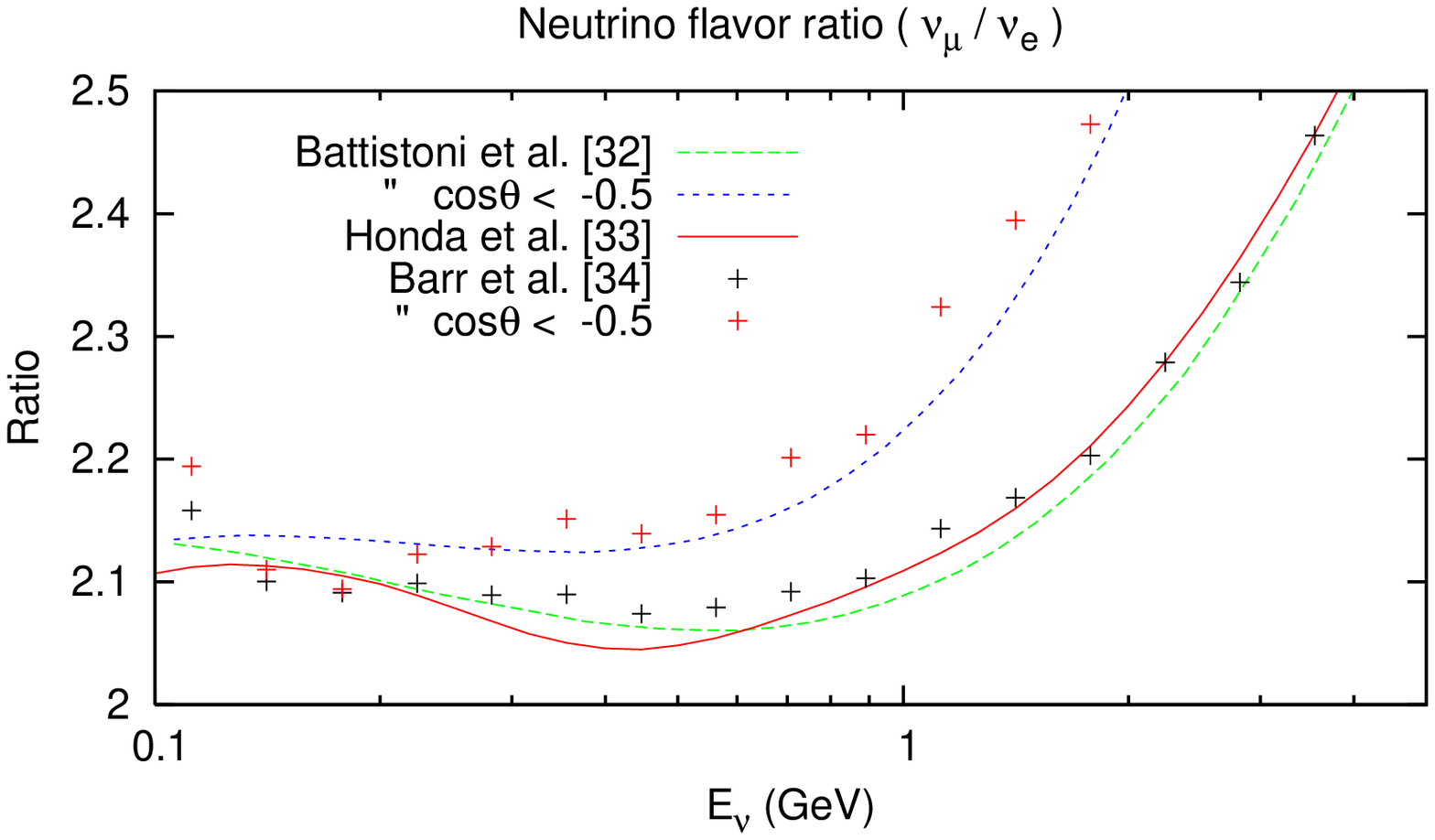}
\caption{\label{fig1b}Comparison of the flavor ratio $r$ 
from three calculations~\protect\cite{FLUKA,Hondaetal,Barretal}.}
\end{minipage} 
\end{figure}

Atmospheric neutrinos are of interest as a beam for the study of neutrino oscillations
and as the background and calibration beam in the search for neutrinos from 
astrophysical sources.  The basic features of the flux of atmospheric neutrinos
have been known since 1961.  Fig.~\ref{fig1a} is a plot of the numerical formulas of
Zatsepin \& Kuz'min~\cite{ZK2}, which shows the main features of
of the flux of atmospheric neutrinos at production.  At low energy there are approximately
two $\nu_\mu+\bar{\nu}_\mu$ produced for each $\nu_e+\bar{\nu}_e$ as a
consequence of the decay sequence,
$$
\pi^\pm\;\rightarrow \;\mu^\pm+\nu_\mu(\bar{\nu}_\mu)\;\rightarrow e^\pm 
+ \nu_e(\bar{\nu}_e) + \bar{\nu}_\mu(\nu_\mu).
$$
The flavor ratio 
\begin{equation}
r\;\equiv\;{\nu_\mu+\bar{\nu}_\mu \over \nu_e+\bar{\nu}_e}
\label{eq1}
\end{equation}
increases with energy above a GeV because muons begin to reach
the ground before they decay.  Some modern calculations of the
muon flavor ratio~\cite{FLUKA,Hondaetal,Barretal} are shown in Fig.~\ref{fig1b}.

The first detections of atmospheric neutrinos were made in the early sixties
in deep mines by Reines {\it et al.} in South Africa~\cite{CW}  and by Menon {\it et al.}
in the Kolar Gold Fields in India~\cite{KGF}.  I have reviewed the history of 
atmospheric neutrino calculations and measurements in more detail elsewhere~\cite{HagaSlott}.
The modern era began in the 1980's with the construction of large underground
detectors to search for proton decay.  Interactions of atmospheric neutrinos
are most numerous in the GeV range and hence constitute the main background for
nucleon decay.  Increasingly precise measurements of the atmospheric neutrino
beam led to the discovery of oscillations~\cite{SK98}
in the $\nu_\mu\leftrightarrow\nu_\tau$ sector, as is well-known.

After a brief discussion of the current level of uncertainties in the flux of atmospheric
neutrinos and the implications for atmospheric neutrinos as a beam for the study of
oscillations, I conclude with some comments on atmospheric neutrinos as background for
searches for astrophysical neutrinos.

\section{Uncertainties in the flux of atmospheric neutrinos}

I want to distinguish three approaches to this subject.  The first is to compare
various calculations of the atmospheric neutrino flux, as in Fig.~\ref{fig1b}.
Other examples of such comparison plots are given in Refs.~\cite{HagaSlott,Neutrino2002,Neutrino2004}.
There is now a large number of calculations that use different approaches,
different interaction models and different representations of the primary 
cosmic-ray spectrum~\cite{FLUKA,Hondaetal,Barretal,nu1,nu2,nu3,nu4}.  
The size of differences among the various calculations can be used to guage
the uncertainty in the neutrino flux.  The general conclusion of such exercises is that
ratios agree to better than 5\% while the uncertainty in normalization
is larger and increases with energy.  Differences among the three calculations
shown in Fig.~\ref{fig1b} for the flavor ratio $r$ are at the level of $2$\%.

A related approach~\cite{SimonBarretal} is to vary the input parameters within the
framework of a single calculational scheme.  This approach seeks to avoid the danger
of different calculations converging on similar results because they use
common input assumptions.  Uncertainties in
hadronic interaction model dominate at low energy, while uncertainties in
the primary spectrum become the dominant source of uncertainty above a few GeV.
Within the set of parameters that characterize uncertainties in hadron production,
those related to production of pions dominate at lower energy, while uncertainties
in strange particle production dominate above 10 GeV, becoming comparable to
the uncertainties from the primary spectrum in the TeV region.  The
importance of kaons is a consequence
of the kinematics of meson decay convolved with a steep primary proton beam, 
which has the effect of making kaon production relatively more important for 
neutrinos than for muons.  For $E_\nu>100$~GeV, kaons become the dominant source
of atmospheric neutrinos.  (See e.g. Fig.~8 of Ref.~\cite{Neutrino2002}).
For analogous reasons, neutrinos from decay of charmed hadrons must eventually become
the most abundant at sufficiently high energy even though charmed hadrons are produced
much less often than strange hadrons.  At some point (e.g. around several hundred TeV),
undertainties in hadro-production of charm will become the biggest source of uncertainty.

\begin{figure}[h]
\begin{minipage}{18pc}
\includegraphics[width=18pc]{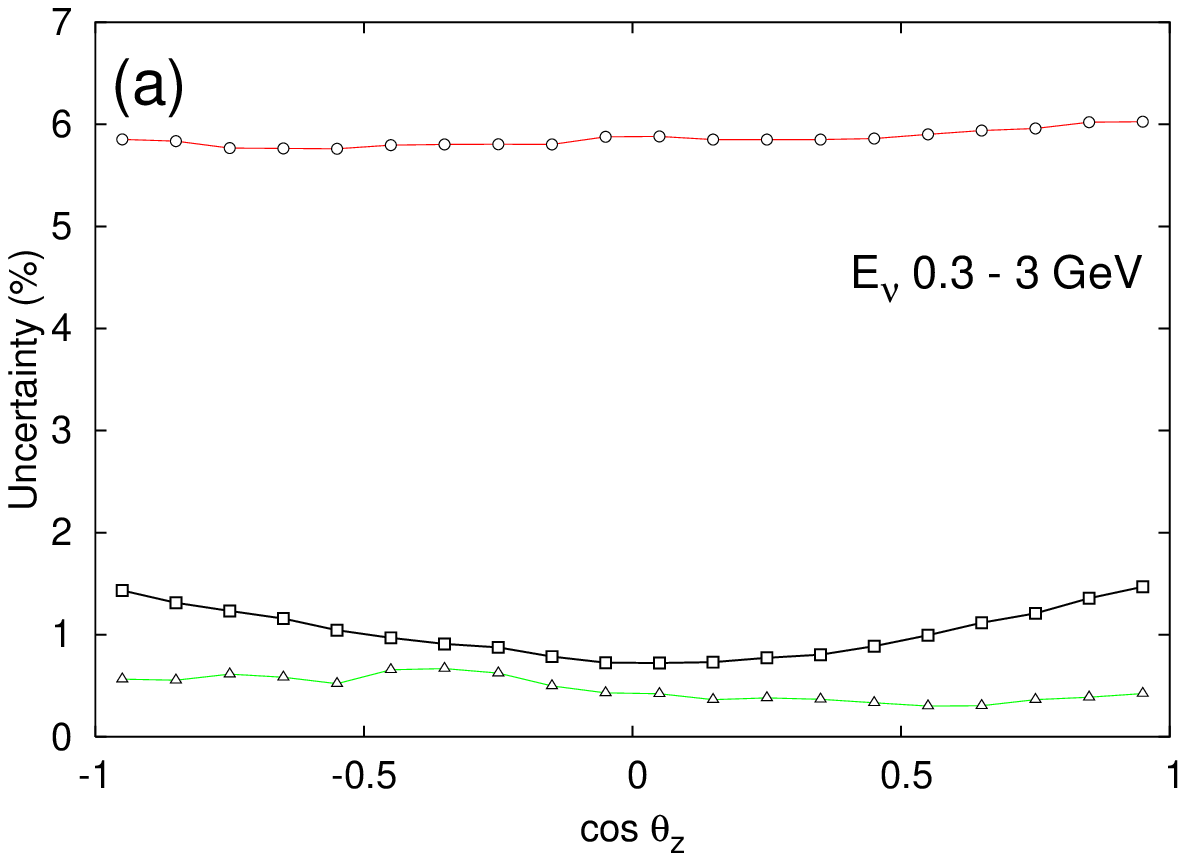}
\caption{\label{fig2a}Uncertainties in neutrino ratios as estimated in
Ref.~\protect\cite{SimonBarretal} ($0.3$-$3$~GeV). (See text.)}
\end{minipage}\hspace{2pc}%
\begin{minipage}{18pc}
\includegraphics[width=18pc]{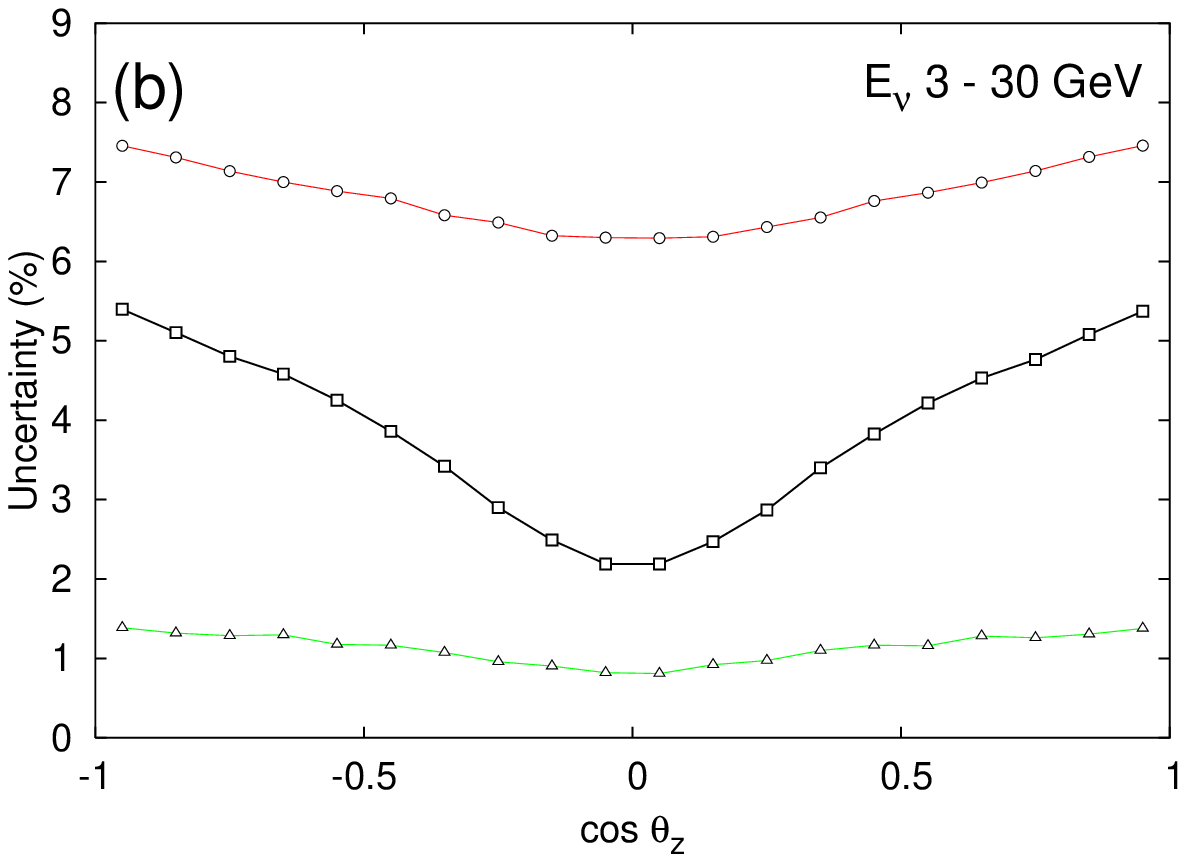}
\caption{\label{fig2b}Same as (a) for $3$-$30$~GeV.  
Lowest curve in (a) and in (b) shows $r$.}
\end{minipage} 
\end{figure}

Overall uncertainty is at the level of $\pm15$\% in the GeV range, rising to $\pm40$\%
for $E_\nu=1$~TeV.  In contrast, uncertainties in the ratios are much smaller
because uncertainties in the primary spectrum and in hadronic interactions cancel
in lowest order in the ratios.  The uncertainty in 
the flavor ratio of Eq.~\ref{eq1} is of order $\pm1$\% for $E_\nu<30$~GeV, as
illustrated in Figs.~\ref{fig2a},\ref{fig2b}.  These figures also show the ratios of neutrinos to
anti-neutrinos, which are somewhat larger than the uncertainty in $r$ 
(6-7\% for $\nu_e/\bar{\nu}_e$ and 1-5\% for $\nu_\mu/\bar{\nu}_\mu$) because
they are more sensitive to the charge ratio of the parent mesons.

A precise knowledge of the flavor ratio $r$ is particularly important in
searching for sub-dominant oscillation effects with atmospheric neutrinos.
For example, oscillations driven by the solar parameters 
are suppressed in the atmospheric neutrino beam by a factor
that depends on the 
near equality of the three neutrino flavors in the 
{\em oscillated} atmospheric neutrino beam~\cite{PerezSmirnov}.  The
observed number of $\nu_e$ ($N_e$) deviates from its value in the absence of 
solar effects by~\cite{PerezSmirnov}
\begin{equation}
{N_e\over N_e(0)} \,-\,1\;=\;P_2 \times(r \, \cos^2\theta_{23}\,-\,1),
\end{equation}  
where $P_2(\delta m^2_{12}$,$\theta_{12}$) is the two-flavor mixing of 
$\nu_e$ with the orthogonal combination of $\nu_\mu$ and $\nu_\tau$~\cite{PerezSmirnov}. 
In the sub-GeV region where pathlengths
comparable to $R_\oplus$ are long enough so that oscillations in the solar
parameters can occur, $r$ is close to two.  Since the atmospheric mixing is
characterized by $\theta_{23}\sim 45^o$ and $\cos^2\theta_{23}\sim 0.5$,
the cancellation is nearly complete.  As shown in Fig.~\ref{fig1b}, however,
$r_{\rm sub-GeV}$ is somewhat larger than two (more so for atmospheric neutrinos in the
vertically upward quadrant of phase space, which have 
pathlength~$>R_\oplus$), making a measurement of the octant of $\theta_{23}$
possible in principle with sufficient statistics.  

A similar suppression factor occurs in the atmospheric neutrino beam for effects
that depend on the deviation of $\sin^2\theta_{13}$ from zero~\cite{Petcov}.
Such effects are, however, expected to be most important for $E_\nu\sim 5$~GeV~\cite{Petcov},
where the flavor ratio is already significantly larger than two.  In this case,
the limiting factor is the intrinsically small size of  $\sin^2\theta_{13}$~\cite{Petcov}.

A different and complementary approach to determining the flux of
atmospheric neutrinos accurately is
to consider the analysis of the data of Super-K~I~\cite{SK1489} 
as a measurement of the flux of atmospheric neutrinos.  The Super-K analysis
proceeds by simultaneously fitting their data with the oscillation parameters together with a large
set of parameters that characterize experimental and theoretical uncertainties.
The theoretical parameters reflect deviations from the assumed
production spectrum of atmospheric neutrinos (i.e. before oscillations).  Shifts in the
fitted parameters that describe the trial production spectrum of atmospheric neutrinos
can be considered as a measurement of the atmospheric neutrino flux at production. 
This approach may be of greatest value for $E_\nu\sim 100$~GeV to $1$~TeV, because the
normalization and oscillation parameters are primarily determined by the
data at lower energy.  In this regard, the adjustment of the spectral index found
in the Super-K fit suggests that the neutrino spectrum continues into the high-energy
range at a higher level than some calculations.  A recent analysis~\cite{concha2} 
confirms this conclusion, as does the new analysis of Honda {\it et al.}~\cite{Honda-new}.  

\begin{figure}[h]
\includegraphics[width=17pc]{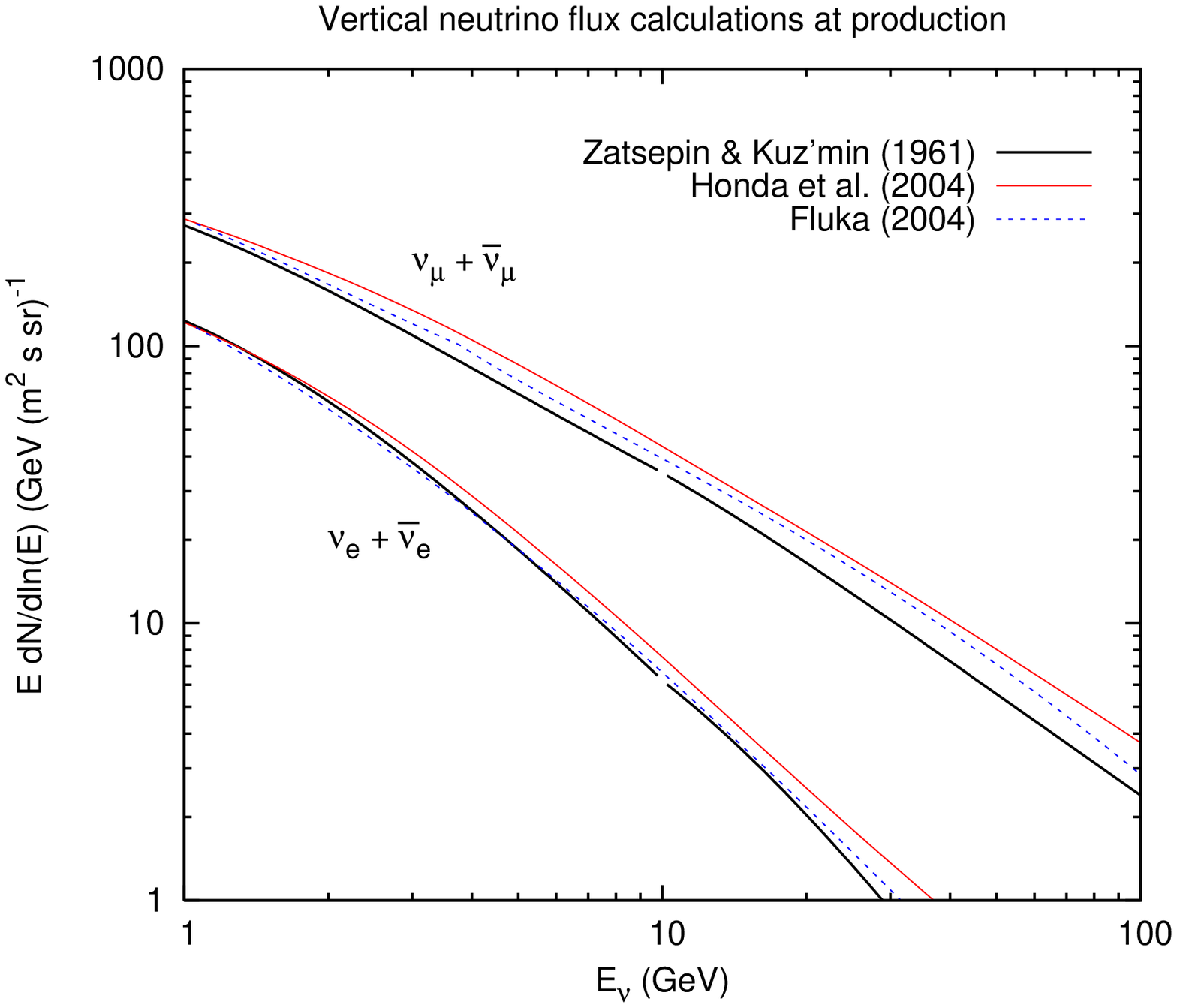}\hspace{2pc}
\includegraphics[width=17pc]{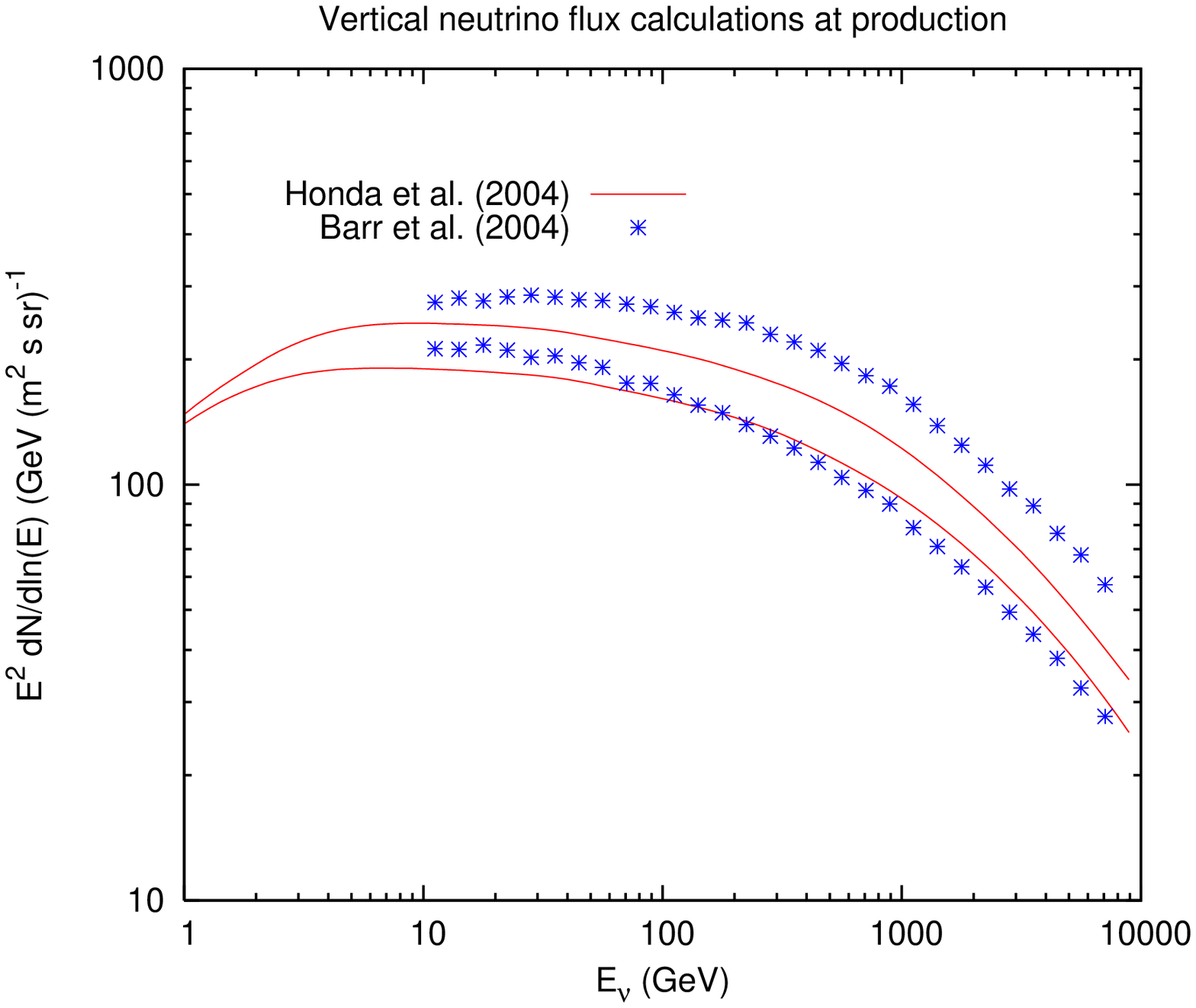}
\caption{\label{fig3}Neutrino flux from several calculations.
The right panel shows muon neutrinos only, with $\nu_\mu$ and $\bar{\nu}_\mu$ 
plotted separately for each calculation.  (Note the
difference in energy ranges and powers of $E$ in the two plots.)
}
\end{figure}

\section{Background for astrophysical neutrinos}

Neutrino telescopes designed to search for astrophysical neutrinos generallly have
thresholds in the range of $\sim 100$~GeV or higher.  Figure~\ref{fig3} is a compilation
of several calculations, incuding two that extend to high energy.  The hard spectral
index that comes out of the Super-K analysis~\cite{SK1489} suggests that higher
intensities should be preferred in the TeV region.  The much larger
ratio of $\nu_\mu\,/\bar{\nu}_\mu$ in Ref.~\cite{Barretal} as compared to
that of Ref.~\cite{Hondaetal} reflects the large associated production 
($p\rightarrow \Lambda\,K^+$) assumed by Barr et al.~\cite{Barretal}
at high energy.  The production of strange and charmed particles
is a significant source of uncertainty and needs more investigation. 
Decay of charmed hadrons
is expected to become the dominant source of atmospheric neutrinos at sufficiently
high energies, $\sim 100$~TeV.  At some level it will become the limiting factor
in a search for a diffuse flux of extra-terrestrial neutrinos.

A well-understood feature of the atmospheric neutrino flux that may be 
useful in distinguishing signal from background is its characteristic
dependence on zenith angle.   A standard form for the differential
spectrum of $\nu_\mu\,+\,\bar{\nu}_\mu$ at high energy is
\begin{equation}
\phi_\nu(E_\nu)\; = \; {\phi_N(E_\nu)\over 1\,-\,Z_{NN}}\times
\Sigma_{i=1}^3\,{A_i\over 1+B_i\cos\theta E_\nu/\epsilon_i},
\end{equation}
where the three terms correspond respectively to neutrinos from
decay of pions, kaons and charmed hadrons.  The overall flux
is proportional to the primary spectrum of nucleons, $\phi_N(E_\nu)$,
evaluated at the energy of the neutrino and scaled by
a factor $1/(1-Z_{NN})$ related to the nucleon attenuation
length.  Each flavor of hadron
has a characteristic critical energy, $\epsilon_i$, above which
the hadron is more likely to interact than to decay.  The
shape of each contribution also depends on a numerical factor ($B_i$)
and on the cosine of the zenith angle.  The latter is the 
``secant theta" effect.  For $E_\nu >> \epsilon_i/(B_i\cos\theta)$
the contribution is inversely proportional to $\cos\theta$ and
asymptotically one power of energy steeper than the primary spectrum.
At very large angles ($\sim \theta > 70^o$) the secant theta
term is limited by the curvature of the Earth.

Neutrinos from astrophysical sources do not depend on the local
zenith angle at which they are observed.  Therefore
in principle the known zenith angle dependence of the atmospheric
background is available as an extra parameter to distinguish
background from signal.  The most obvious example would be the
contrast between atmospheric background and an isotropic, diffuse
extraterrestrial flux of high-energy neutrinos.  Because the
contribution from charm is also isotropic (until extremely high
energy), the distinction disappears when the intensity of the
extraterrestrial neutrinos is at the level of atmospheric neutrinos
from decay of charmed hadrons.

\begin{figure}[h]
\begin{minipage}{18pc}
\includegraphics[width=18pc]{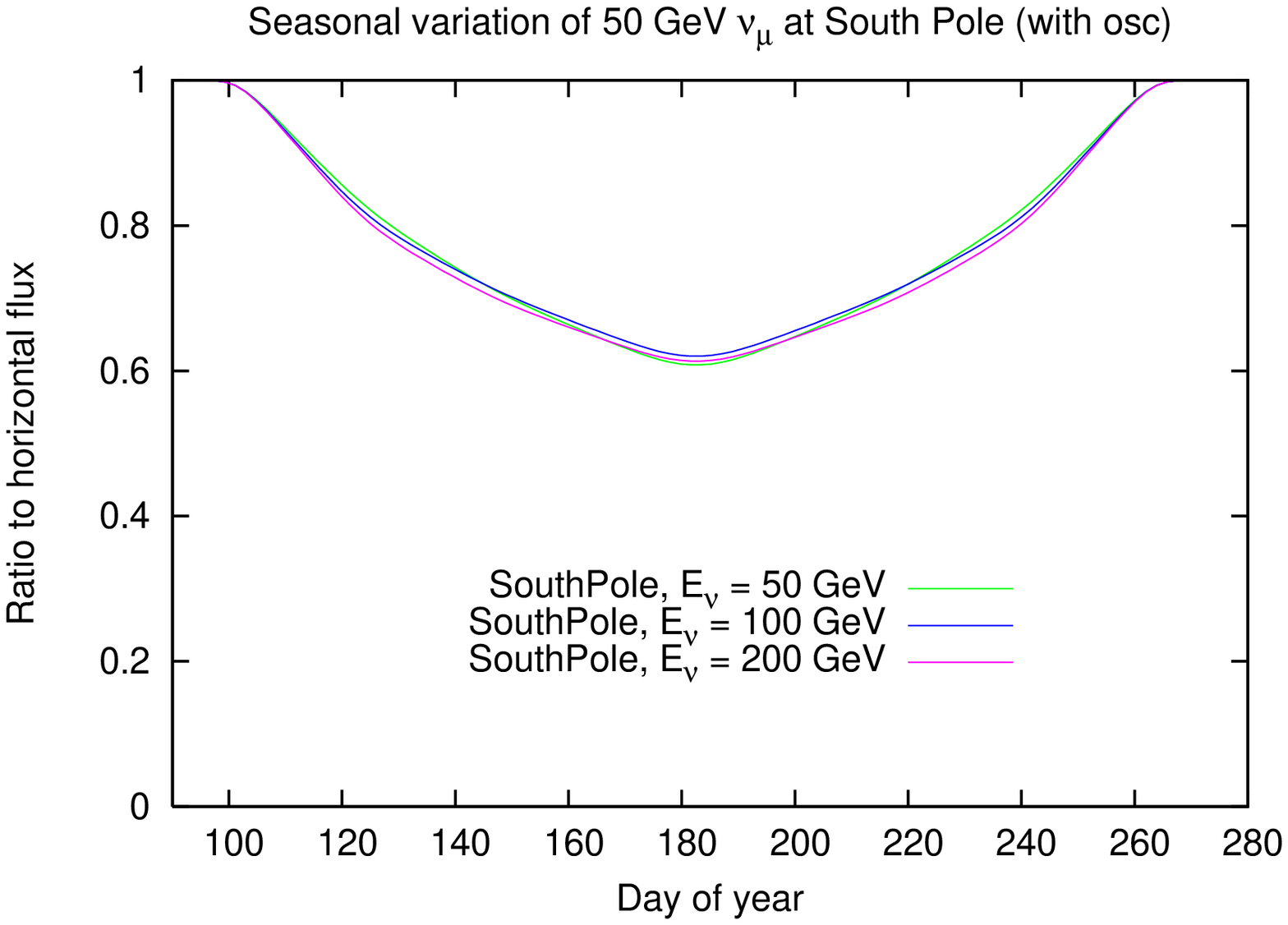}
\caption{\label{fig4a}Relative variation of the intensity of
atmospheric $\nu_\mu +\bar{\nu}_\mu$
from the direction of the Sun as viewed from the South Pole
during Austral winter (including oscillations).  
}
\end{minipage}\hspace{2pc}%
\begin{minipage}{18pc}
\includegraphics[width=18pc]{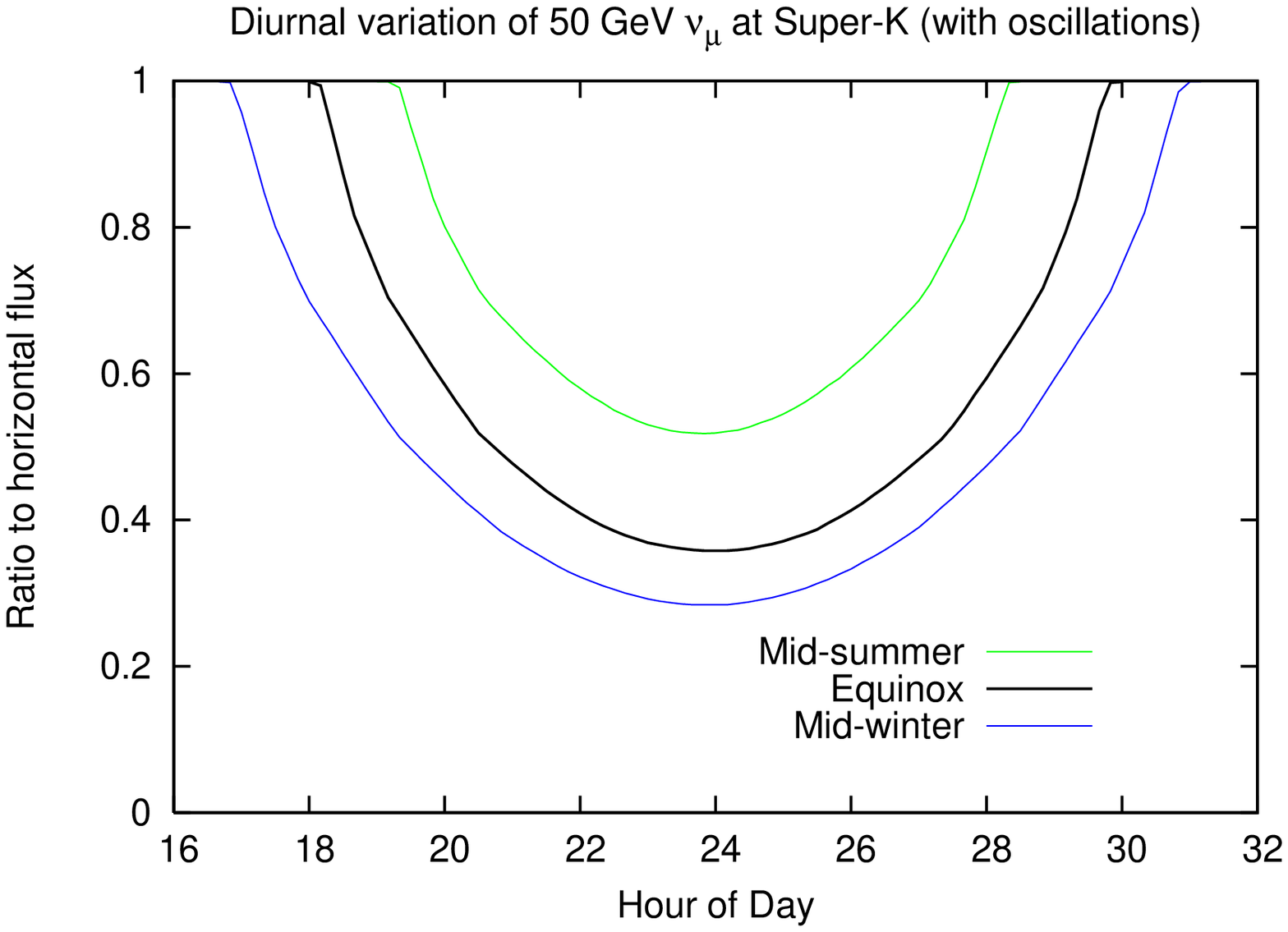}
\caption{\label{fig4b}  Relative variation of the intensity
of $50$~GeV atmospheric $\nu_\mu +\bar{\nu}_\mu$
from the direction of the Sun as viewed from Super-K,
including oscillation effects.}
\end{minipage} 
\end{figure}

For point sources of neutrinos observed from mid-latitude detectors, 
variation of the background in the direction of a potential source
as it rises and sets can in principle also help to distinguish
background from signal.  A related example is the indirect search
for neutrinos from WIMP annihilation in the Sun.  Figures~\ref{fig4a}\ref{fig4b}
show the expected diurnal variation of the atmsopheric background
from the direction of the Sun as seen from the South Pole
and from Super-K.

\end{document}